\documentclass{ifacconf}

\usepackage[table]{xcolor}
\usepackage{graphicx}
\usepackage{natbib}
\usepackage{amsmath,amsfonts}
\usepackage{bm}
\usepackage{nicematrix}
\usepackage{subcaption}
\DeclareMathOperator*{\st}{subject~to}

\DeclareMathOperator*{\argmin}{arg\,min}

\newcommand{\norm}[1]{\left\lVert#1\right\rVert}
\definecolor{darkgreen}{rgb}{0.0, 0.4, 0.0}
\definecolor{lightgray}{gray}{0.9}
\makeatletter
\let\old@ssect\@ssect 
\makeatother

\usepackage{hyperref}

\makeatletter
\def\@ssect#1#2#3#4#5#6{%
  \NR@gettitle{#6}
  \old@ssect{#1}{#2}{#3}{#4}{#5}{#6}
}
\makeatother

\usepackage{balance}
\begin{document}
\begin{frontmatter}

\title{Follow the Clairvoyant: an Imitation Learning Approach to Optimal Control\thanksref{footnoteinfo}}

\thanks[footnoteinfo]{Research supported by the Swiss National Science Foundation under the NCCR Automation (grant agreement 51NF40\textunderscore180545).}

\author[First]{Andrea Martin}
\author[First]{Luca Furieri}
\author[Second]{Florian D\"orfler}
\author[Second]{John Lygeros}
\author[First]{Giancarlo Ferrari-Trecate}

\address[First]{Institute of Mechanical Engineering, \'Ecole Polytechnique F\'ed\'erale de Lausanne, Switzerland (e-mail: \{andrea.martin, luca.furieri, giancarlo.ferraritrecate\}@epfl.ch).}
\address[Second]{Department of Information Technology and Electrical Engineering, ETH Z\"urich, Switzerland (e-mail: \{dorfler, jlygeros\}@ethz.ch)}

\begin{abstract}
    We consider control of dynamical systems through the lens of competitive analysis. Most prior work in this area focuses on minimizing regret, that is, the loss relative to an ideal clairvoyant policy that has noncausal access to past, present, and future disturbances. Motivated by the observation that the optimal cost only provides coarse information about the ideal closed-loop behavior, we instead propose directly minimizing the tracking error relative to the optimal trajectories in hindsight, i.e., imitating the clairvoyant policy. By embracing a system level perspective, we present an efficient optimization-based approach for computing follow-the-clairvoyant (FTC) safe controllers. We prove that these attain minimal regret if no constraints are imposed on the noncausal benchmark. In addition, we present numerical experiments to show that our policy retains the hallmark of competitive algorithms of interpolating between classical $\mathcal{H}_2$ and $\mathcal{H}_\infty$ control laws – while consistently outperforming regret minimization methods in constrained scenarios thanks to the superior ability to chase the clairvoyant.
\end{abstract}


\end{frontmatter}

\section{Introduction}
Motivated by online learning methods, optimal control of uncertain dynamical systems has recently been studied from the modern perspective of regret minimization \citep{goel2021regret, sabag2021regret}. In this competitive paradigm, control laws are designed to minimize the maximal loss relative to an ideal clairvoyant policy that has access to past, present, and future disturbances. As such, regret-optimal policies offer tight performance guarantees in terms of the cost incurred by an optimal policy in hindsight, which hold independently of the stochastic or adversarial nature of the disturbances. This property stands in stark contrast with the distributional and worst-case assumptions that characterize classical $\mathcal{H}_2$ and $\mathcal{H}_\infty$ controllers \citep{hassibi1999indefinite}, and makes regret an attractive control design criterion.

By leveraging operator-theoretic techniques, analytical solutions to the regret-optimal control problem have first been derived in \cite{goel2021regret} and \cite{sabag2021regret}. By embracing a system level approach \citep{wang2019system}, these methods have then been extended to account for convex safety constraints in \cite{martin2022safe} and \cite{didier2022system}, whereas the related metric of competitive ratio has been considered in \cite{goel2021competitive} and \cite{sabag2022optimal}. 

Unlike these works and inspired by imitation learning (IL) approaches \citep{hussein2017imitation, yin2021imitation}, in this paper we view the clairvoyant policy as an expert providing demonstrations of desired behavior, and we propose designing causal controllers that track these optimal trajectories – not merely the cost that they attain. Intuitively, this problem is more challenging, as there may be multiple trajectories that result in the same control cost. 
On the other hand, aggregate measures such as the control cost are generally less informative, as they only describe the ideal behavior in a parsimonious way. Therefore, granting novice controllers access to the entire expert trajectories may improve the overall performance, as it provides them with more insights on how the optimal cost has been achieved. Besides, this formulation appears more natural in constrained scenarios, where safety imposes limits on the physical variables of the system while, at the same time, efficiency demands operation close to the boundaries of the safe set \citep{mayne2000constrained}. In these settings, following as closely as possible a safe trajectory that has been computed with foreknowledge of the unknowns could in fact prove key for steering the system towards the edges of the feasible set, and for achieving near-optimal performance in the face of uncertain disturbance realizations.

The contributions of this paper can be summarized as follows. First, we introduce a novel competitive metric that is reminiscent of behavioral cloning approaches in the IL literature \citep[Chapter 3]{osa2018algorithmic}. Leveraging the system level parametrization of linear dynamic controllers \citep{wang2019system}, we show that the minimization of the worst-case tracking error relative to the trajectory of a clairvoyant policy can be equivalently formulated as a semidefinite optimization program, which can be efficiently solved via convex optimization techniques. Second, we reveal strong connections with the regret-optimal control framework of \cite{goel2021regret} and \cite{sabag2021regret}. In particular, we prove that, if one chooses the unique noncausal optimal policy as the unconstrained control benchmark \citep{hassibi1999indefinite}, then FTC policies attain minimal regret and, vice-versa, regret-optimal controllers incur minimal imitation loss. We hope that our theoretical insights will stimulate the development of novel competitive policy design methods for real-world applications such as mobile robotics, where generating causal reference trajectories currently stands as a major computational bottleneck \citep{romero2022model}. As soon as more complex benchmarks are considered, however, the equivalence between tracking optimal trajectories and minimizing regret ceases to hold. Last, we present numerical experiments to illustrate how the superior ability of our policy to chase the clairvoyant can yield improved closed-loop performance in constrained scenarios.

\section{Problem Statement and Preliminaries}
We consider a discrete-time linear time-varying dynamical system described by the state-space equation
\begin{equation}
    \label{eq:system_dynamics}
    x_{t+1} = A_t x_t + B_t u_t + w_t\,,
\end{equation}
where $x_t \in \mathbb{R}^n$, $u_t \in \mathbb{R}^m$, and $w_t \in \mathbb{R}^n$ are the system state, the control input, and an exogenous disturbance process, respectively. We study the evolution of this system over a finite time horizon of length $T \in \mathbb{N}$, and we use the compact notation $\mathbf{x} = (x_0, \dots, x_{T-1})$, $\mathbf{u} = (u_0, \dots, u_{T-1})$, and $\mathbf{w} = (x_0, w_0, \dots, w_{T-2})$. At each sampling instant, a measurement of the state $x_t$ becomes available, and a control action $u_t = \pi_t(x_0, x_1, \dots, x_t)$ is computed by the online controller according to a causal decision policy $\bm{\pi} = (\pi_0, \dots, \pi_{T-1})$. Along the control horizon, the controller incurs the quadratic cost
\begin{equation}
    \label{eq:lqr_cost}
    J(\bm{\pi}, \mathbf{w}) = \mathbf{x}^\top \mathbf{Q} \mathbf{x} + \mathbf{u}^\top \mathbf{R} \mathbf{u}\,,
\end{equation}
where $\mathbf{Q} \succeq 0$ and $\mathbf{R} \succ 0$ are the state and input cost matrices, respectively. Since \eqref{eq:lqr_cost} depends on yet unknown disturbance realizations, computing the cost-minimizing sequence of control actions given causal information becomes an ill-posed problem. To break this deadlock, classical $\mathcal{H}_2$ and $\mathcal{H}_\infty$ control theories pose specific disturbance-generating mechanisms and minimize the expected and the worst-case value of \eqref{eq:lqr_cost}, respectively. Conversely, towards designing adaptive control laws that adjust to the true disturbance realizations, regret minimization algorithms \citep{goel2021regret, sabag2021regret} do not make any assumptions about the nature of the disturbance process. Instead, these methods compute an optimal controller by solving the minimax competitive problem
\begin{equation}
    \label{eq:regret_minimization_definition}
    \min_{\bm{\pi}} ~ \max_{\norm{\mathbf{w}}_2 \leq 1} ~ \left[J(\bm{\pi}, \mathbf{w}) - J(\bm{\psi}, \mathbf{w})\right]\,,
\end{equation}
where $\bm{\psi}$ denotes a given clairvoyant feedback policy, which plays the role of a control benchmark.\footnote{Often times, $\bm{\psi}$ is selected as the unique unconstrained clairvoyant optimal policy that minimizes \eqref{eq:lqr_cost} globally \citep{hassibi1999indefinite}; we will reserve the symbol $\bm{\psi}^\star$ to refer to this specific unconstrained benchmark, and instead use $\bm{\psi}$ to denote an arbitrary, possibly constrained, clairvoyant policy that is linear in $\mathbf{w}$, see \cite{martin2022safe}.} Here, inspired by behavioral cloning approaches in IL, we instead propose designing an optimal policy by solving the minimax tracking problem
\begin{equation}
    \label{eq:imitation_loss_minimization_definition}
    \min_{\bm{\pi}} ~ \max_{\norm{\mathbf{w}}_2 \leq 1} ~ \left[\bm{\delta}_{x, \psi}^\top \mathbf{Q} \bm{\delta}_{x, \psi} + 
    \bm{\delta}_{u, \psi}^\top \mathbf{R} 
    \bm{\delta}_{u, \psi}\right]\,,
\end{equation}
where $\bm{\delta}_{x, \psi} = \mathbf{x} - \mathbf{x}_\psi$, $\bm{\delta}_{u, \psi} = \mathbf{u} - \mathbf{u}_\psi$, and $\mathbf{x}_\psi$ and $\mathbf{u}_\psi$ denote the expert state and input trajectories under an appropriately designed clairvoyant policy $\bm{\psi}$. We remark that, by leveraging our knowledge of the system dynamics \eqref{eq:system_dynamics} and the cost \eqref{eq:lqr_cost}, we can explicitly compute the clairvoyant reference trajectories $\mathbf{x}_\psi$ and $\mathbf{u}_\psi$ by proceeding as in Corollary~4 of \cite{martin2022safe}. The main challenge of imitating the expert behavior thus lies in staying close to the optimal trajectory in hindsight given causal information only.\footnote{This stands in contrast with the challenge, typical of imitation learning \citep{hussein2017imitation, osa2018algorithmic}, of learning from a finite number of demonstrations in unknown environments.} The difference between regret minimization and FTC methods is the following: by penalizing the difference between the incurred costs, \eqref{eq:regret_minimization_definition} only uses the cost $J(\bm{\psi}, \mathbf{w})$ as aggregate information about the optimal closed-loop trajectory, whereas \eqref{eq:imitation_loss_minimization_definition} takes full advantage of the knowledge about the expert policy $\bm{\psi}$.

\begin{rem}
    \label{rem:weights}
    The weights in the imitation loss \eqref{eq:imitation_loss_minimization_definition} need not be equal to the cost matrices in \eqref{eq:lqr_cost}, and may be selected, for instance, to prioritize tracking of the expert trajectories $\mathbf{x}_\psi$ and $\mathbf{u}_\psi$ along particular state and input components. For ease of presentation, however, in this paper we will not exploit this greater design flexibility.
\end{rem}

Since many control applications require fulfilling state and input constraints, we consider the objective of minimizing the worst-case imitation loss while guaranteeing that the system operates safely. As dynamic programming solutions are generally computationally intractable, we restrict our attention to linear feedback policies $\mathbf{u} = \mathbf{K} \mathbf{x}$, with $\mathbf{K}$ lower block-triangular to enforce causality. Then, we consider a polytopic safe set
\begin{equation}
    \label{eq:safe_set_definition}
    {\mathcal{S} = \{(\mathbf{x}, \mathbf{u}) : \mathbf{H}_x \mathbf{x} + \mathbf{H}_u \mathbf{u} \leq \mathbf{h}\}\,,\footnotemark}
\end{equation}
\footnotetext{Inequalities involving vectors apply element-wise.}and require the novice policy $\bm{\pi}$ to ensure that $(\mathbf{x}, \mathbf{u}) \in \mathcal{S}$ robustly for all disturbance sequences
\begin{equation}
    \label{eq:disturbance_set_definition}
    \mathbf{w} \in \mathcal{W} = \{\mathbf{w} : \mathbf{H}_w \mathbf{w} \leq \mathbf{h}_w\}\,,
\end{equation}
that belong to the compact polytope $\mathcal{W}$. As common in the robust predictive control literature \citep{mayne2000constrained, rawlings2017model}, we assume that there exists a controller $\mathbf{K}$ that complies with the state and input constraints for all possible $\mathbf{w} \in \mathcal{W}$. If this were not the case, one would need to consider richer classes of policies, or to relax the safety requirements, for instance by introducing slack variables in \eqref{eq:safe_set_definition}. Finally, we remark that satisfaction of this feasibility assumption can be checked by solving a linear optimization problem, see also the reformulation of the safety constraints by duality arguments in Theorem~\ref{th:ftc_sdp}.

We conclude our problem formulation by reviewing useful results on the unconstrained clairvoyant optimal policy $\bm{\psi}^\star$ that minimizes \eqref{eq:lqr_cost} globally \citep{hassibi1999indefinite}. The synthesis of FTC controllers crucially relies on the availability of a tractable description of the expert policy. In principle, obtaining such a concise representation could prove more challenging than only appraising the cost that the expert incurs by means of oracle evaluations. Fortunately, this is not the case. In particular, we now show that $\bm{\psi}^\star$ simply performs a linear combination of past, present and future realizations of the disturbance. 

Let the solution of the linear dynamics \eqref{eq:system_dynamics} be encoded as $\mathbf{x} = \mathbf{F}\mathbf{u} + \mathbf{G} \mathbf{w}$, where $\mathbf{w}$ encloses the unknown system initial condition $x_0 \in \mathbb{R}^n$, and $\mathbf{F}$ and $\mathbf{G}$ are causal response operators comprising the system Markov parameters obtained by recursion of \eqref{eq:system_dynamics}. With this notation in place, we first rewrite the quadratic loss \eqref{eq:lqr_cost} as
\begin{align*}
    J(\bm{\pi}, \mathbf{w})
    &= \left(\mathbf{F} \mathbf{u} + \mathbf{G} \mathbf{w} \right)^\top \mathbf{Q} \left(\mathbf{F} \mathbf{u} + \mathbf{G} \mathbf{w} \right) + \mathbf{u}^\top \mathbf{R} \mathbf{u}\\
    &= \mathbf{u}^\top \mathbf{P} \mathbf{u} + 2 \mathbf{u}^\top \mathbf{F}^\top \mathbf{Q} \mathbf{G} \mathbf{w} + \mathbf{w}^\top \mathbf{G}^\top \mathbf{Q} \mathbf{G} \mathbf{w} \,,
\end{align*}
where $\mathbf{P} = \mathbf{R} + \mathbf{F}^\top \mathbf{Q}\mathbf{F} \succ 0$. 
Let $\mathbf{I}$ denote an identity matrix of appropriate dimensions, and observe that the identity
\begin{equation*}
    \mathbf{Q}\mathbf{F}\mathbf{P}^{-1}\mathbf{F}^\top \mathbf{Q} + \mathbf{Q}(\mathbf{I} + \mathbf{F}\mathbf{R}^{-1}\mathbf{F}^\top\mathbf{Q})^{-1} = \mathbf{Q}
\end{equation*}
holds by the matrix inversion lemma. Then, we equivalently express the incurred control cost as
\begin{align}
    \label{eq:lqr_cost_split}
    J(\bm{\pi}, \mathbf{w}) &= (\mathbf{P}\mathbf{u} + \mathbf{F}^\top \mathbf{Q} \mathbf{G} \mathbf{w})^\top \mathbf{P}^{-1} (\mathbf{P}\mathbf{u} + \mathbf{F}^\top \mathbf{Q} \mathbf{G} \mathbf{w}) + \nonumber\\
    &\quad+ \mathbf{w}^\top \mathbf{G}^\top \mathbf{Q} (\mathbf{I} + \mathbf{F}\mathbf{R}^{-1}\mathbf{F}^\top \mathbf{Q})^{-1} \mathbf{G} \mathbf{w}\,,
\end{align}
namely, as the sum of a non-negative quantity and a term that only depends on the realizations of the disturbance – not on the choice of the control inputs. In absence of safety constraints, the globally optimal clairvoyant policy $\bm{\psi}^\star$ can hence be obtained by simply setting the former term equal to zero, which directly yields
\begin{subequations}
    \begin{align}
        \label{eq:clairvoyant_optimal_policy_definition}
        \mathbf{u}_{\psi^\star}(\mathbf{w}) &= -(\mathbf{R} + \mathbf{F}^\top\mathbf{Q} \mathbf{F})^{-1}\mathbf{F}^\top\mathbf{Q}\mathbf{G}\mathbf{w}\,,\\
        \label{eq:clairvoyant_optimal_policy_cost}
        J(\bm{\psi}^\star, \mathbf{w}) &= \mathbf{w}^\top \mathbf{G}^\top \mathbf{Q}(\mathbf{I} + \mathbf{F}\mathbf{R}^{-1}\mathbf{F}^\top\mathbf{Q})^{-1} \mathbf{G} \mathbf{w}\,.
    \end{align}
\end{subequations}
By construction, \eqref{eq:clairvoyant_optimal_policy_cost} represents a lower bound on the cost incurred by any possibly nonlinear clairvoyant policy.

\section{Main Results}
In this section, we present our main results. First, we present a tractable optimization-based approach for computing novice policies that safely minimize the imitation loss in \eqref{eq:imitation_loss_minimization_definition}; we refer to these as FTC policies. Second, we reveal connections with the regret minimization framework of \cite{goel2021regret} and \cite{sabag2021regret} by showing that – independently of the novice policy $\bm{\pi}$ being subject to safety requirements or not – FTC policies achieve minimal regret if no constraints are imposed on the noncausal benchmark, i.e., if $\bm{\psi} = \bm{\psi}^\star$. To do so, we leverage the system level parametrization of linear dynamic controllers \citep{wang2019system}, which we briefly review hereafter, to shift the synthesis problem from directly designing the controller to shaping the closed-loop maps from the exogenous disturbance to the state and input signals. 

Let $\mathbf{Z}$ be the block-downshift operator, i.e., a matrix with identity matrices along its first block sub-diagonal and zeros elsewhere, $\mathbf{A} = \operatorname{blkdiag}(A_0, A_1, \dots, A_{T-2}, 0_{n \times n})$, and $\mathbf{B} = \operatorname{blkdiag}(B_0, B_1, \dots, B_{T-2}, 0_{n \times m})$. Further, note that $\mathbf{F} = (\mathbf{I} - \mathbf{Z} \mathbf{A})^{-1} \mathbf{Z} \mathbf{B}$ and $\mathbf{G} = (\mathbf{I} - \mathbf{Z} \mathbf{A})^{-1}$. We define the closed-loop system responses $\bm{\Phi}_x$ and $\bm{\Phi}_u$ as
\begin{subequations}
    \label{eq:causal_closed_loop_responses_definition}
    \begin{align}
        \mathbf{x} &= (\mathbf{I} - \mathbf{Z}(\mathbf{A} + \mathbf{B} \mathbf{K}))^{-1} \mathbf{w} = \bm{\Phi}_x \mathbf{w}\,,\\
        \mathbf{u} &= \mathbf{K} \mathbf{x} = \mathbf{K} (\mathbf{I} - \mathbf{Z}(\mathbf{A} + \mathbf{B} \mathbf{K}))^{-1} \mathbf{w} = \bm{\Phi}_u \mathbf{w} \label{eq:causal_closed_loop_responses_definition_u}\,.
    \end{align}
\end{subequations}
Then, we recall that the linear achievability constraint 
\begin{equation}
    \label{eq:sls_achievability}
    (\mathbf{I} - \mathbf{Z}\mathbf{A})\bm{\Phi}_x - \mathbf{Z}\mathbf{B} \bm{\Phi}_u = \mathbf{I}\,,
\end{equation}
provides a necessary and sufficient condition for the existence of a linear feedback controller $\mathbf{K} = \bm{\Phi}_u \bm{\Phi}_x^{-1}$ such that $\mathbf{x} = \bm{\Phi}_x \mathbf{w}$ and $\mathbf{u} = \bm{\Phi}_u \mathbf{w}$, that is, such that the desired closed-loop system responses are achievable \citep{wang2019system}. Furthermore, we remark that the system responses constructed by \eqref{eq:causal_closed_loop_responses_definition} inherit a lower block-triangular structure from the causal sparsity of the controller $\mathbf{K}$. As shown in \cite{martin2022safe}, the optimal clairvoyant policy $\bm{\psi}^\star$ characterized in \eqref{eq:clairvoyant_optimal_policy_definition} can also be computed by directly optimizing over the set of noncausal system responses that are not necessarily lower block-triangular. Further, this optimization-oriented perspective naturally allows one to include safety requirements and to define more complex benchmark policies. Given a possibly constrained clairvoyant policy $\bm{\psi}$, we denote by $\bm{\Psi}_x$ and $\bm{\Psi}_u$ the noncausal system responses that achieve the corresponding closed-loop behavior, i.e.,  $\mathbf{x}_\psi = \bm{\Psi}_x \mathbf{w}$ and $\mathbf{u}_\psi = \bm{\Psi}_u \mathbf{w}$; by inspection of \eqref{eq:clairvoyant_optimal_policy_definition}, we also note that $\bm{\Psi}_u^\star = -(\mathbf{R} + \mathbf{F}^\top\mathbf{Q} \mathbf{F})^{-1}\mathbf{F}^\top\mathbf{Q}\mathbf{G}$.

For compactness, we define $\mathbf{C} = \operatorname{blkdiag}(\mathbf{Q}, \mathbf{R})$ and $\mathbf{H} = \begin{bmatrix} \mathbf{H}_x & \mathbf{H}_u\end{bmatrix}$ through diagonal and horizontal concatenation of matrices, respectively. We are now ready to derive a convex reformulation of the FTC synthesis problem.
\begin{thm}
    \label{th:ftc_sdp}
    Let the linear time-varying system \eqref{eq:system_dynamics} evolve over a control horizon of length $T \in \mathbb{N}$, and consider the constrained FTC optimal control problem:
    \begin{subequations}
        \label{eq:ftc_formulation}
        \begin{align}
            & ~ \min_{\bm{\pi}} ~ \max_{\norm{\mathbf{w}}_2 \leq 1} ~ \left[\bm{\delta}_{x, \psi}^\top \mathbf{Q} \bm{\delta}_{x, \psi} + \bm{\delta}_{u, \psi}^\top \mathbf{R} \bm{\delta}_{u, \psi}\right]\\
            &\st \quad (\mathbf{x}, \mathbf{u}) \in \mathcal{S}\,, ~ \forall \mathbf{w} \in \mathcal{W}\,,
        \end{align}  
    \end{subequations}
    defined with respect to the possibly constrained clairvoyant linear policy $\bm{\psi}$. Then, \eqref{eq:ftc_formulation} admits the following equivalent formulation as semidefinite optimization problem:
    \begin{subequations}
        \label{eq:ftc_sdp}
        \begin{align}
            & ~ \min_{\bm{\Phi}_x, \bm{\Phi}_u, \mathbf{Y}, \lambda} ~ \lambda\\
            & \st ~ \eqref{eq:sls_achievability}\,, ~ \lambda \geq 0\,, \nonumber\\
            & \qquad
            \mathbf{Y}^\top \mathbf{h}_w \leq \mathbf{h}\,, ~
            \mathbf{H}
            \begin{bmatrix}
                \bm{\Phi}_x\\\bm{\Phi}_u
            \end{bmatrix} = \mathbf{Y}^\top \mathbf{H}_w\,, ~ \mathbf{Y}_{ij} \geq 0\,,\label{eq:ftc_dual_safety_constraints}\\
            & \qquad \begin{bmatrix}
                \mathbf{I} & 
                \mathbf{C}^{\frac{1}{2}}
                \begin{bmatrix}
                    \bm{\Phi}_x - \bm{\Psi}_x\\
                    \bm{\Phi}_u - \bm{\Psi}_u\\
                \end{bmatrix}
                \\
                \begin{bmatrix}
                    \bm{\Phi}_x - \bm{\Psi}_x\\
                    \bm{\Phi}_u - \bm{\Psi}_u\\
                \end{bmatrix}^\top
                \mathbf{C}^{\frac{1}{2}} & \lambda \mathbf{I}
            \end{bmatrix} \succeq 0\,, \label{eq:sdp_schur_complement}\\
            & \qquad \bm{\Phi}_x, \bm{\Phi}_u \text{ with causal sparsities,} \label{eq:system_responses_sparisites}
        \end{align}
    \end{subequations}
    where $\bm{\Psi}_x$ and $\bm{\Psi}_u$ are the noncausal system responses that correspond to the benchmark policy $\bm{\psi}$.
\end{thm}
\textit{Proof. } 
By definition, the imitation loss \eqref{eq:imitation_loss_minimization_definition} compares the trajectories of the novice and expert policies driven by the same disturbance. From \eqref{eq:causal_closed_loop_responses_definition}, we then have that $\bm{\delta}_{x,\psi} = (\bm{\Phi}_x - \bm{\Psi}_x) \mathbf{w}$ and $\bm{\delta}_{u,\psi} = (\bm{\Phi}_u - \bm{\Psi}_u) \mathbf{w}$. Hence, \eqref{eq:ftc_formulation} can be equivalently expressed as
\begin{subequations}
    \begin{align}
        & ~ \min_{\bm{\Phi}_x, \bm{\Phi}_u}  ~ \max_{\|\mathbf{w}\|_{2} \leq 1}
        ~ \mathbf{w}^\top 
        \underbrace{
        \begin{bmatrix}
            \bm{\Phi}_x - \bm{\Psi}_x\\
            \bm{\Phi}_u - \bm{\Psi}_u
        \end{bmatrix}^\top
        \mathbf{C}
        \begin{bmatrix}
            \bm{\Phi}_x - \bm{\Psi}_x\\
            \bm{\Phi}_u - \bm{\Psi}_u
        \end{bmatrix}
        }_{\bm{\Delta}(\bm{\Phi}_x, \bm{\Phi}_u, \bm{\Psi}_x, \bm{\Psi}_u)}
        \mathbf{w} \label{eq:ftc_minimax_delta} \\
        & \st ~\eqref{eq:sls_achievability}, \eqref{eq:system_responses_sparisites}\nonumber\,, \\
        & \qquad \mathbf{H}
        \begin{bmatrix}
            \bm{\Phi}_x\\
            \bm{\Phi}_u
        \end{bmatrix} \mathbf{w} \leq \mathbf{h}\,, ~ \forall \mathbf{w} \text{ such that } \mathbf{H}_w \mathbf{w} \leq \mathbf{h}_w \label{eq:ftc_safety_constraints_forall}\,.
    \end{align}  
\end{subequations}
To derive a tractable reformulation of \eqref{eq:ftc_minimax_delta}, we first note that, by construction and independently of $\bm{\Psi}_x$ and $\bm{\Psi}_u$, $\bm{\Delta} \succeq 0$ for all choices of $\bm{\Phi}_x$ and $\bm{\Phi}_u$.\footnote{When clear from the context, we omit function arguments in the interest of readability.} Hence, denoting by $\lambda_{\operatorname{max}}(\cdot)$ the maximum eigenvalue of a matrix, we have that
\begin{align*}
    \max_{\|\mathbf{w}\|_{2} \leq 1} ~ \mathbf{w}^\top \bm{\Delta} \mathbf{w} = \lambda_{\operatorname{max}}(\bm{\Delta}) = & ~ \min_{\lambda \geq 0} ~ \lambda\\ 
    & \st ~ \lambda \mathbf{I} - \bm{\Delta} \succeq 0\,,
\end{align*}
where the last equality follows from classical results on semidefinite programming for eigenvalue minimization (see, for instance, Section 2.2 in \cite{boyd1994linear}). Then, as $\bm{\Delta}$ still depends quadratically on the optimization variables $\bm{\Phi}_x$ and $\bm{\Phi}_u$, we apply the Schur complement, which yields the linear matrix inequality constraint \eqref{eq:sdp_schur_complement}. 

To ensure that the safety constraints are robustly satisfied, we proceed as in \cite{martin2022safe} and employ strong duality of linear optimization problems to eliminate the universal quantifier from \eqref{eq:ftc_safety_constraints_forall}. Specifically, we note that 
\begin{alignat*}{3}
    \max_{\mathbf{w} \in \mathcal{W}} ~ \left(
    \mathbf{H}
    \begin{bmatrix}
        \bm{\Phi}_x\\
        \bm{\Phi}_u
    \end{bmatrix}\right)_i \mathbf{w} = & ~ \min_{\mathbf{y}_i \geq 0} ~ \mathbf{h}_w^\top \mathbf{y}_i\,,\\ 
    & \st ~ \mathbf{H}_w^\top \mathbf{y}_i = \left(
    \mathbf{H}
    \begin{bmatrix}
        \bm{\Phi}_x\\
        \bm{\Phi}_u
    \end{bmatrix}
    \right)_i^\top\,, 
\end{alignat*}
where $\mathbf{y}_i$ denotes the dual vector corresponding with the $i$-th row of the maximization. From this dual reformulation, the set of linear constraints \eqref{eq:ftc_dual_safety_constraints} can be derived by concatenating the dual variables that arise from each row of the maximization into the matrix $\mathbf{Y}$.
$\hfill \qed$

Note that an optimal solution $\{\bm{\Phi}_x^\star, \bm{\Phi}_u^\star\}$ to \eqref{eq:ftc_sdp} can be efficiently computed thanks to convexity, and that the corresponding optimal state-feedback controller can then be reconstructed by $\mathbf{K}^\star = \bm{\Phi}_u^\star {\bm{\Phi}_x^\star}^{-1}$. Independently of how the disturbances are generated, this optimal control policy guarantees minimum worst-case tracking error relative to the optimal trajectories of the clairvoyant policy $\bm{\psi}$. 

As we will illustrate by means of numerical simulations, this property proves particularly valuable in constrained scenarios, where, due to limited actuation power and safety requirements, the optimal cost only carries coarse information about the ideal closed-loop behavior. On the other hand, one could reasonably expect this property to be of interest even in unconstrained scenarios, as closely following the optimal trajectory in hindsight should result in near-optimal performance. Our next result formalizes this intuition by showing that – no matter how tight the safety constraints imposed on the novice policy $\bm{\pi}$ are – the objectives of \eqref{eq:regret_minimization_definition} and \eqref{eq:imitation_loss_minimization_definition} become equivalent if $\bm{\psi}$ is chosen as the unconstrained clairvoyant optimal policy $\bm{\psi}^\star$.

\begin{thm}
    \label{th:ftc_reg}
    Assume that no constraints are imposed on the clairvoyant benchmark policy, that is, let $\bm{\psi} = \bm{\psi}^\star$. Then, for any linear control policy $\bm{\pi}$,
    \begin{equation}
    \label{eq:connection_regret_imitiation_loss}
        J(\bm{\pi}, \mathbf{w}) - J(\bm{\psi}^\star, \mathbf{w}) = \bm{\delta}_{x, \psi^\star}^\top \mathbf{Q} \bm{\delta}_{x, \psi^\star} + 
    \bm{\delta}_{u, \psi^\star}^\top \mathbf{R} 
    \bm{\delta}_{u, \psi^\star}\,,
    \end{equation} 
    that is, the regret equals the imitation loss for all possible disturbance realizations.
\end{thm}
\textit{Proof. }
By combining \eqref{eq:lqr_cost_split} with \eqref{eq:clairvoyant_optimal_policy_cost}, we first express the left-hand side of \eqref{eq:connection_regret_imitiation_loss} as
\begin{equation*}
    (\mathbf{P} \mathbf{u} + \mathbf{F}^\top \mathbf{Q} \mathbf{G} \mathbf{w})^\top \mathbf{P}^{-1} (\mathbf{P} \mathbf{u} + \mathbf{F}^\top \mathbf{Q} \mathbf{G} \mathbf{w})\,.
\end{equation*}
Then, observing that $\mathbf{P}^{-1} = \mathbf{P}^{-1} \mathbf{P} \mathbf{P}^{-1}$ and letting $\mathbf{u} = \bm{\Phi}_u \mathbf{w}$ as in \eqref{eq:causal_closed_loop_responses_definition_u}, the regret becomes
\begin{equation}
    \label{eq:proof_th3_1}
    (\bm{\Phi}_u \mathbf{w} + \mathbf{P}^{-1} \mathbf{F}^\top \mathbf{Q} \mathbf{G} \mathbf{w})^\top \mathbf{P} (\bm{\Phi}_u \mathbf{w} + \mathbf{P}^{-1} \mathbf{F}^\top \mathbf{Q} \mathbf{G} \mathbf{w})\,.
\end{equation}
Recalling that $\mathbf{P} = \mathbf{R} + \mathbf{F}^\top \mathbf{Q} \mathbf{F}$ by definition and using \eqref{eq:clairvoyant_optimal_policy_definition}, we now recognize that $\mathbf{P}^{-1} \mathbf{F}^\top \mathbf{Q} \mathbf{G} = - \bm{\Psi}_u^\star$, where $\bm{\Psi}_u^\star$ is the unconstrained noncausal system response that maps the disturbances $\mathbf{w}$ to $\mathbf{u}_{\psi^\star}$, the optimal sequence of control inputs in hindsight. Rearranging terms, we can equivalently rewrite \eqref{eq:proof_th3_1} as
\begin{align}
    &\mathbf{w}^\top (\bm{\Phi}_u - \bm{\Psi}_u^\star)^\top (\mathbf{R} + \mathbf{F}^\top \mathbf{Q} \mathbf{F}) (\bm{\Phi}_u - \bm{\Psi}_u^\star) \mathbf{w} \nonumber \\
    & \quad \quad = \mathbf{w}^\top 
    \begin{bmatrix}
        \mathbf{F} \bm{\Phi}_u - \mathbf{F} \bm{\Psi}_u^\star\\
        \bm{\Phi}_u - \bm{\Psi}_u^\star
    \end{bmatrix}^\top
    \mathbf{C}
    \begin{bmatrix}
        \mathbf{F} \bm{\Phi}_u - \mathbf{F} \bm{\Psi}_u^\star\\
        \bm{\Phi}_u - \bm{\Psi}_u^\star
    \end{bmatrix}
    \mathbf{w}\,. \label{eq:proof_th3_2}
\end{align}
Finally, we note that achievability constraint \eqref{eq:sls_achievability} implies that the closed-loop responses associated with $\bm{\pi}$ and $\bm{\psi}^\star$ satisfy the relations
\begin{align*}
    \bm{\Phi}_x &= (\mathbf{I} - \mathbf{Z} \mathbf{A})^{-1} \mathbf{Z} \mathbf{B} \bm{\Phi}_u + (\mathbf{I} - \mathbf{Z} \mathbf{A})^{-1} = \mathbf{F} \bm{\Phi}_u + \mathbf{G}\,,\\ 
    \bm{\Psi}_x^\star &= (\mathbf{I} - \mathbf{Z} \mathbf{A})^{-1} \mathbf{Z} \mathbf{B} \bm{\Psi}_u^\star + (\mathbf{I} - \mathbf{Z} \mathbf{A})^{-1} = \mathbf{F} \bm{\Psi}_u^\star + \mathbf{G}\,.
\end{align*}
Hence, $\mathbf{F} \bm{\Phi}_u - \mathbf{F} \bm{\Psi}_u^\star = \bm{\Phi}_x - \bm{\Psi}_x^\star$, and the proof is concluded by comparing \eqref{eq:proof_th3_2} with \eqref{eq:ftc_minimax_delta}.
$\hfill \qed$

If $\bm{\psi} = \bm{\psi}^\star$, the equivalence established in Theorem~\ref{th:ftc_reg} guarantees that by minimizing the imitation loss \eqref{eq:imitation_loss_minimization_definition}, one simultaneously minimizes the regret \eqref{eq:regret_minimization_definition}, and – perhaps more surprisingly as multiple trajectories could yield the same control cost – that also the converse implication holds. As we show next, besides bridging two different competitive metrics, this relation also provides a novel characterization of the classical $\mathcal{H}_2$ controller as the decision policy that minimizes the expected imitation loss relative to the clairvoyant optimal policy $\bm{\psi}^\star$.
\begin{cor}
    Let $\bm{\psi} = \bm{\psi}^\star$ and assume that $\mathbf{w}$ follows a probability distribution $\mathcal{D}$ with mean $\bm{\mu}_\mathbf{w}$ and covariance matrix $\Sigma_\mathbf{w} \succeq 0$, that is, $\mathbf{w} \sim \mathcal{D}(\bm{\mu}_\mathbf{w}, \Sigma_\mathbf{w})$. Then, the set of minimizers of $\mathbb{E}[\bm{\delta}_{x, \psi^\star}^\top \mathbf{Q} \bm{\delta}_{x, \psi^\star} + \bm{\delta}_{u, \psi^\star}^\top \mathbf{R} \bm{\delta}_{u, \psi^\star}]$ coincides with $\argmin_{\bm{\pi}} ~ \mathbb{E}[J(\bm{\pi}, \mathbf{w})]$.
\end{cor}
\textit{Proof. }
Recall that \eqref{eq:connection_regret_imitiation_loss} holds for every $\mathbf{w}$ by reason of Theorem~\ref{th:ftc_reg}. Then, taking weighted averages, we have that
\begin{align*}
    & \mathbb{E}[\bm{\delta}_{x, \psi^\star}^\top \mathbf{Q} \bm{\delta}_{x, \psi^\star} + 
    \bm{\delta}_{u, \psi^\star}^\top \mathbf{R} 
    \bm{\delta}_{u, \psi^\star}]\\
    & = \mathbb{E}[J(\bm{\pi}, \mathbf{w}) - J(\bm{\psi}^\star, \mathbf{w})]
    = \mathbb{E}[J(\bm{\pi}, \mathbf{w})] - \mathbb{E}[J(\bm{\psi}^\star, \mathbf{w})]\,,
\end{align*}
where the second equality follows from linearity of the expectation. By observing that $\mathbb{E}[J(\bm{\psi}^\star, \mathbf{w})]$ does not depend on the control policy $\bm{\pi}$, we directly conclude that minimizing the expected imitation loss is equivalent to minimizing the expected control cost $\mathbb{E}[J(\bm{\pi}, \mathbf{w})]$. 
$\hfill \qed$

We highlight that the equivalence between minimizing regret and tracking clairvoyant optimal trajectories breaks as soon as we consider more complex constrained benchmarks, or if we select different weights for the control cost \eqref{eq:lqr_cost} and the imitation loss \eqref{eq:imitation_loss_minimization_definition}, see also Remark~\ref{rem:weights}. In the next section, we compare the behavior of regret-optimal and FTC policies by means of numerical simulations. 

\section{Numerical Results}
We now validate numerically the equivalence proved in Theorem~\ref{th:ftc_reg}, and we then show how imitating the clairvoyant policy can lead to improved performance in constrained scenarios. For our experiments, we consider the open-loop unstable system \eqref{eq:system_dynamics} with
\begin{equation*}
    A_t = \rho \begin{bmatrix}
        0.7 & 0.2 & 0\\
        0.3 & 0.7 & -0.1\\
        0 & -0.2 & 0.8
    \end{bmatrix}\,, ~
    B_t = \begin{bmatrix}
        1 & 0.2\\
        2 & 0.3\\
        1.5 & 0.5
    \end{bmatrix}\,, ~ \forall t \in \mathbb{I}_{T}\,,
\end{equation*}
where $\rho = 1.05$ is the spectral radius of the system, $\mathbb{I}_{T} = \{0, \dots T-1\}$, and the control horizon is $T = 30$. We define the control cost \eqref{eq:lqr_cost} and the imitation loss \eqref{eq:imitation_loss_minimization_definition} by letting $\mathbf{Q} = \mathbf{I}_{30} \otimes \mathbf{I}_{3}$ and $\mathbf{R} = \mathbf{I}_{30} \otimes \mathbf{I}_{2}$, where $\otimes$ denotes the Kronecker product. Further, we consider the safe set
\begin{equation*}
    \mathcal{S} = \{(\mathbf{x}, \mathbf{u}) : -10 \leq x_{t} \leq 10\,, ~ -10 \leq u_t \leq 10\,, ~ \forall t \in \mathbb{I}_{T}\}\,,
\end{equation*}
and enforce robust constraints satisfaction for all initial conditions and disturbance realizations taking values in
\begin{equation*}
    \mathcal{W} = \{\mathbf{w} : -1 \leq x_0 \leq 1\,, -1 \leq w_t \leq 1\,, \forall t \in \mathbb{I}_{T}\}\,.
\end{equation*}
We first compute $\bm{\psi}$ as the clairvoyant $\mathcal{H}_2$ safe policy by leveraging the results in Corollary~4 of \cite{martin2022safe}. Then, we solve the semidefinite optimization problem \eqref{eq:ftc_sdp} to synthesize a FTC policy $\bm{\pi}$ that minimizes the worst-case imitation loss while complying with the prescribed safety requirements. By inspection, we observe that the chosen clairvoyant benchmark does not activate the safety constraints, namely, that $\bm{\psi} = \bm{\psi}^\star$. Therefore, as predicted by Theorem~\eqref{th:ftc_reg}, we verify that $\bm{\pi}$ achieves optimal regret and, in particular, that the maximum loss relative to the globally optimal sequence of control actions in hindsight is $\max_{\norm{\mathbf{w}}_2 \leq 1} ~ \left[J(\bm{\pi}, \mathbf{w}) - J(\bm{\psi}, \mathbf{w})\right] = 7.98$. We then compare the average costs incurred by our FTC policy and by classical $\mathcal{H}_2$ and $\mathcal{H}_\infty$ safe controllers for a variety of stochastic and deterministic disturbances, see Table~\ref{table:results}.\footnote{The code that reproduces our numerical examples is available at \href{https://github.com/DecodEPFL/FollowTheClairvoyant}{https://github.com/DecodEPFL/FollowTheClairvoyant}. Please refer to the simulation code for a precise definition of the disturbance profiles that appear in Table~\ref{table:results}.} As expected, our numerical results closely mirror those obtained in \cite{martin2022safe}. This shows that FTC policies inherit from regret minimization methods the capability to interpolate between the performance of $\mathcal{H}_2$ and $\mathcal{H}_\infty$ control laws, often outperforming them when the true disturbances do not match classical design assumptions.

\begin{table}[ht]
    \captionsetup{width=\columnwidth}
    \caption{Average control cost increase relative\\to the control policy denoted with {\color{darkgreen} \textbf{1}}.}
    \label{table:results}
    \centering
    \rowcolors{1}{}{lightgray}
    \begin{tabular}{c|cccc}
        \hline
        $\mathbf{w}$ & $\mathcal{H}_2$ & $\mathcal{H}_\infty$ & $\operatorname{FTC}$\\
        \hline
        $\mathcal{N}(0,1)$ & \color{darkgreen} \textbf{1} & $>$+100\% & +52.33\%\\
        $\mathcal{U}_{[0.5, 1]}$ & +37.79\% & +10.86\% & \color{darkgreen} \textbf{1}\\
        $\mathcal{U}_{[0, 1]}$ & +5.90\% & +20.48\% & \color{darkgreen} \textbf{1}\\
        $1$ & +46.00\% & +8.31\% & \color{darkgreen} \textbf{1}\\
        $\operatorname{sin}$ & +38.36\% & +12.29\% & \color{darkgreen} \textbf{1}\\
        $\operatorname{sawtooth}$ & +29.74\% & +16.41\% & \color{darkgreen} \textbf{1}\\
        $\operatorname{step}$ & +15.66\% & \color{darkgreen} \textbf{1} & +0.51\%\\
        $\operatorname{stairs}$ & +18.98\% & +3.60\% & \color{darkgreen} \textbf{1}\\
        $\operatorname{worst}$ & $>$+100\% & \color{darkgreen} \textbf{1} & +26.08\%\\
        \hline
    \end{tabular}
\end{table}

To illustrate how chasing the clairvoyant can improve the overall performance, we tighten the safety constraints and require that the closed-loop input and state trajectories robustly remain inside the safe set
\begin{equation*}
    \mathcal{S} = \{(\mathbf{x}, \mathbf{u}) : -10 \leq x_{t} \leq 10\,, ~ -5 \leq u_t \leq 5\,, ~ \forall t \in \mathbb{I}_{T}\}\,.
\end{equation*}

\begin{figure}[htb]
    \begin{subfigure}{\columnwidth}
      \centering
      \vspace{5pt}
      \includegraphics[width=\columnwidth]{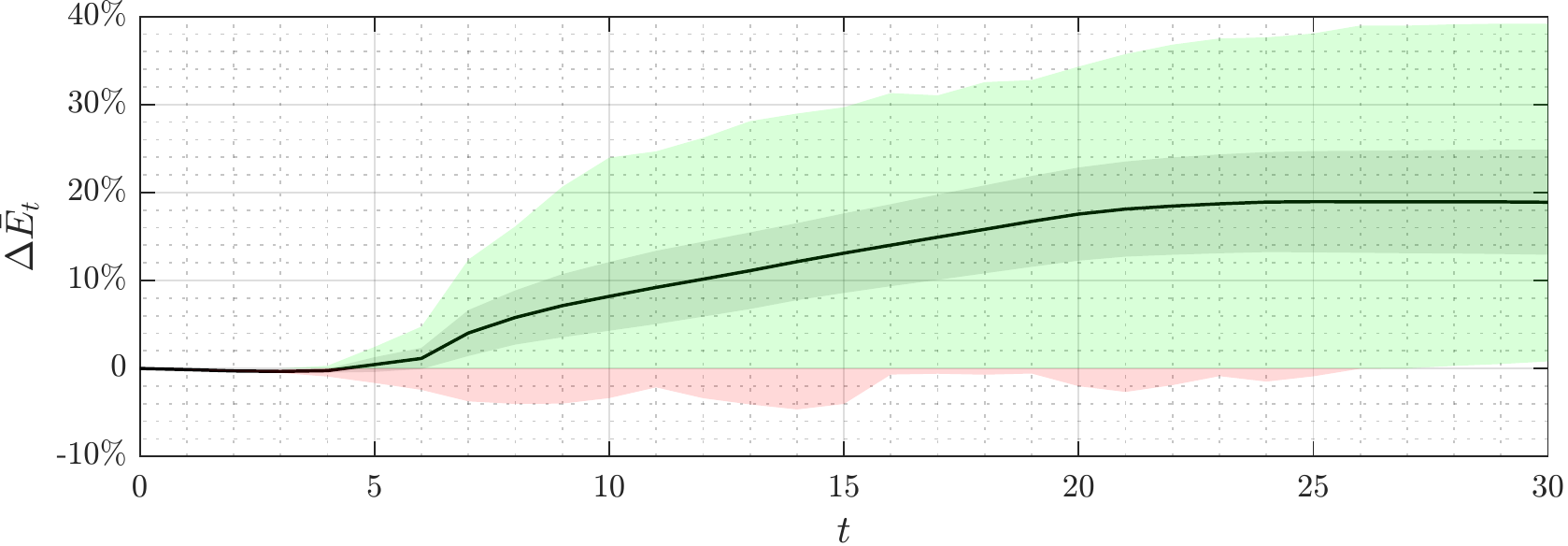}  
      \caption{Tracking error percentage increase relative to $\bm{\pi}_{\operatorname{F}}$.}
      \label{fig:tracking}
    \end{subfigure}
    \vspace{10pt}\\
    \begin{subfigure}{\columnwidth}
      \centering
      \includegraphics[width=\columnwidth]{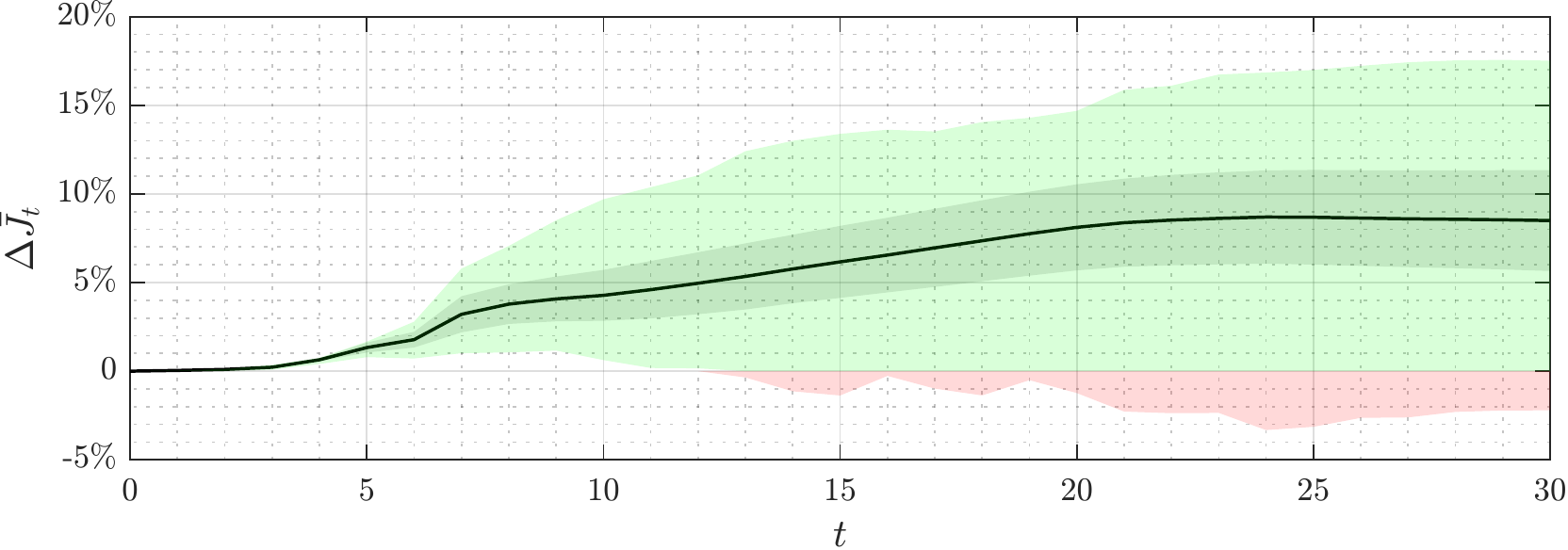}  
      \caption{Control cost percentage increase relative to $\bm{\pi}_{\operatorname{F}}$.}
      \label{fig:cost}
    \end{subfigure}
    \caption{Performance comparison between $\bm{\pi}_{\operatorname{R}}$ and $\bm{\pi}_{\operatorname{F}}$: the darker tubes show the standard deviation around the black continuous line that denotes the average performance difference. Green and red shaded areas delimit the regions containing all realizations of the performance increase relative to $\bm{\pi}_{\operatorname{F}}$.}
\end{figure}

For this second experiment, we compute both a regret-optimal and a FTC safe control policy, which we denote by $\bm{\pi}_{\operatorname{R}}$ and $\bm{\pi}_{\operatorname{F}}$, respectively, by minimizing the loss and the tracking error relative to a clairvoyant $\mathcal{H}_2$ safe policy. We allow the benchmark to select each control action with foreknowledge of the next 5 disturbance realizations; note that, by only granting the benchmark policy a limited preview of the unknowns, we forgo a certain degree of optimality of the clairvoyant policy in return for more causally explainable expert decisions, which can be tracked more consistently.\footnote{This benchmark policy can be computed by restricting the number of non-zero entries in the upper triangular part of the clairvoyant system responses $\bm{\Psi}_x$ and $\bm{\Psi}_u$ in Corollary~4 of \cite{martin2022safe}.} To validate the intuition that our FTC policy $\bm{\pi}_{\operatorname{F}}$ better tracks the optimal trajectories in hindsight close to the edges of the feasible set, we randomly sample from the vertices of the polytope $\mathcal{W}$ a set of disturbance realizations $\mathbf{w}_i$, $i \in \mathbb{I}_N$ with $N = 1000$, that nearly activate the safety constraints under the clairvoyant benchmark policy. Then, for both $\bm{\pi} = \bm{\pi}_{\operatorname{F}}$ and $\bm{\pi} = \bm{\pi}_{\operatorname{R}}$, we compute the average tracking error $\bar{E}_t(\bm{\pi})$ and the average control cost $\bar{J}_t(\bm{\pi})$ that each decision policy incurs up to time $t$ as follows:
\begin{equation*}
    \bar{E}_t(\bm{\pi}) = \frac{1}{N} \sum_{i = 0}^{N-1} E_t(\bm{\pi}, \mathbf{w}_i)\,, ~ 
    \bar{J}_t(\bm{\pi}) = \frac{1}{N} \sum_{i = 0}^{N-1} J_t(\bm{\pi}, \mathbf{w}_i)\,,
\end{equation*}
for all $t \in \mathbb{I}_{T}$, where we define $E_t(\bm{\pi}, \mathbf{w}_i)$ and $J_t(\bm{\pi}, \mathbf{w}_i)$ as
\begin{align*}
    E_t(\bm{\pi}, \mathbf{w}_i) &= \sum_{k = 0}^{t} \delta_{x_k, \psi}^\top \delta_{x_k, \psi} + \delta_{u_k, \psi}^\top \delta_{u_k, \psi}\,, ~ \forall t \in \mathbb{I}_{T}\,,\\
    J_t(\bm{\pi}, \mathbf{w}_i) &= \sum_{k = 0}^{t} x_k^\top x_k + u_k^\top u_k\,, ~ \forall t \in \mathbb{I}_{T}\,,
\end{align*}
and where $\delta_{x_t, \psi}$ and $\delta_{u_t, \psi}$ are the time $t$ components of $\bm{\delta}_{x, \psi}$ and $\bm{\delta}_{u, \psi}$, respectively. Finally, to compare the performance of regret-optimal and FTC policies, we study the percentage difference of the average tracking error and control cost relative to $\bm{\pi}_{\operatorname{F}}$, that is, we evaluate the quantities
\begin{equation*}
    \Delta \bar{E}_t = \frac{\bar{E}_t(\bm{\pi}_{\operatorname{R}}) - \bar{E}_t(\bm{\pi}_{F})}{\bar{E}_t(\bm{\pi}_{F})}\,, ~
    \Delta \bar{J}_t = \frac{\bar{J}_t(\bm{\pi}_{\operatorname{R}}) - \bar{J}_t(\bm{\pi}_{F})}{\bar{J}_t(\bm{\pi}_{F})}\,,
\end{equation*}
for all $t \in \mathbb{I}_T$. We plot the evolution of these two performance metrics over the horizon with a black continuous line in Figure~\ref{fig:tracking} and Figure~\ref{fig:cost}, respectively.
For completeness, we visualize the standard deviation of $\Delta \bar{E}_t$ and $\Delta \bar{J}_t$ by means of two gray shaded tubes around the corresponding mean values. Further, we use green and red shaded areas to denote the regions, associated with positive and negative values on the vertical axis, that contain all the realized trajectories of $E_t(\bm{\pi}_{\operatorname{R}}, \mathbf{w}_i) - E_t(\bm{\pi}_{\operatorname{F}}, \mathbf{w}_i)$ and $J_t(\bm{\pi}_{\operatorname{R}}, \mathbf{w}_i) - J_t(\bm{\pi}_{\operatorname{F}}, \mathbf{w}_i)$, normalized by the constants $\bar{E}_t(\bm{\pi}_{\operatorname{F}})$ and $\bar{J}_t(\bm{\pi}_{\operatorname{F}})$, respectively. Figures~\ref{fig:tracking} and~\ref{fig:cost} show that our FTC safe policy yields closed-loop trajectories that stay consistently closer to the optimal trajectories in hindsight, while the regret-optimal control policy $\bm{\pi}_{\operatorname{R}}$ incurs almost $20\%$ higher imitation loss on average. In turn, we observe that, by following more tightly the ideal closed-loop behavior, the proposed control law often outperforms the regret-optimal control policy even in the regret sense; in this particularly challenging scenario, $\bm{\pi}_{\operatorname{R}}$ pays, on average, almost a $10\%$ higher control cost.

\section{Conclusion}
Inspired by imitation learning approaches, we have introduced a novel competitive metric for designing control laws that minimize the tracking error relative to the optimal closed-loop trajectories in hindsight. We have formally shown the equivalence between following the clairvoyant and minimizing regret if no constraints are imposed on the noncausal benchmark policy. For the more general case of constrained linear control benchmarks, instead, we have presented a framework for synthesizing safe FTC policies efficiently via convex optimization, and we have illustrated the potential of the proposed paradigm by means of numerical simulations. Future research directions include addressing infinite-horizon control problems, developing model-free solutions, and studying the interplay between different competitive metrics for nonlinear systems.

\balance

\bibliography{ifacconf}

\begin{thebibliography}{15}
\providecommand{\natexlab}[1]{#1}
\providecommand{\url}[1]{\texttt{#1}}
\providecommand{\urlprefix}{URL }
\expandafter\ifx\csname urlstyle\endcsname\relax
  \providecommand{\doi}[1]{doi:\discretionary{}{}{}#1}\else
  \providecommand{\doi}{doi:\discretionary{}{}{}\begingroup
  \urlstyle{rm}\Url}\fi

\bibitem[{Boyd et~al.(1994)Boyd, El~Ghaoui, Feron, and
  Balakrishnan}]{boyd1994linear}
Boyd, S., El~Ghaoui, L., Feron, E., and Balakrishnan, V. (1994).
\newblock \emph{Linear matrix inequalities in system and control theory}.
\newblock SIAM.

\bibitem[{Didier et~al.(2022)Didier, Sieber, and Zeilinger}]{didier2022system}
Didier, A., Sieber, J., and Zeilinger, M.N. (2022).
\newblock A system level approach to regret optimal control.
\newblock \emph{IEEE Control Systems Letters}.

\bibitem[{Goel and Hassibi(2021{\natexlab{a}})}]{goel2021competitive}
Goel, G. and Hassibi, B. (2021{\natexlab{a}}).
\newblock Competitive control.
\newblock \emph{arXiv preprint arXiv:2107.13657}.

\bibitem[{Goel and Hassibi(2021{\natexlab{b}})}]{goel2021regret}
Goel, G. and Hassibi, B. (2021{\natexlab{b}}).
\newblock Regret-optimal estimation and control.
\newblock \emph{arXiv preprint arXiv:2106.12097}.

\bibitem[{Hassibi et~al.(1999)Hassibi, Sayed, and
  Kailath}]{hassibi1999indefinite}
Hassibi, B., Sayed, A.H., and Kailath, T. (1999).
\newblock \emph{Indefinite-quadratic estimation and control: a unified approach
  to $\mathcal{H}_2$ and $\mathcal{H}_\infty$ theories}.
\newblock SIAM.

\bibitem[{Hussein et~al.(2017)Hussein, Gaber, Elyan, and
  Jayne}]{hussein2017imitation}
Hussein, A., Gaber, M.M., Elyan, E., and Jayne, C. (2017).
\newblock Imitation learning: A survey of learning methods.
\newblock \emph{ACM Computing Surveys (CSUR)}, 50(2), 1--35.

\bibitem[{Martin et~al.(2022)Martin, Furieri, D{\"o}rfler, Lygeros, and
  Ferrari-Trecate}]{martin2022safe}
Martin, A., Furieri, L., D{\"o}rfler, F., Lygeros, J., and Ferrari-Trecate, G.
  (2022).
\newblock Safe control with minimal regret.
\newblock In \emph{Learning for Dynamics and Control Conference}, 726--738.
  PMLR.

\bibitem[{Mayne et~al.(2000)Mayne, Rawlings, Rao, and
  Scokaert}]{mayne2000constrained}
Mayne, D.Q., Rawlings, J.B., Rao, C.V., and Scokaert, P.O. (2000).
\newblock Constrained model predictive control: Stability and optimality.
\newblock \emph{Automatica}, 36(6), 789--814.

\bibitem[{Osa et~al.(2018)Osa, Pajarinen, Neumann, Bagnell, Abbeel, Peters
  et~al.}]{osa2018algorithmic}
Osa, T., Pajarinen, J., Neumann, G., Bagnell, J.A., Abbeel, P., Peters, J.,
  et~al. (2018).
\newblock An algorithmic perspective on imitation learning.
\newblock \emph{Foundations and Trends{\textregistered} in Robotics}, 7(1-2),
  1--179.

\bibitem[{Rawlings et~al.(2017)Rawlings, Mayne, and Diehl}]{rawlings2017model}
Rawlings, J.B., Mayne, D.Q., and Diehl, M. (2017).
\newblock \emph{Model predictive control: theory, computation, and design},
  volume~2.
\newblock Nob Hill Publishing Madison.

\bibitem[{Romero et~al.(2022)Romero, Sun, Foehn, and
  Scaramuzza}]{romero2022model}
Romero, A., Sun, S., Foehn, P., and Scaramuzza, D. (2022).
\newblock Model predictive contouring control for time-optimal quadrotor
  flight.
\newblock \emph{IEEE Transactions on Robotics}.

\bibitem[{Sabag et~al.(2021)Sabag, Goel, Lale, and Hassibi}]{sabag2021regret}
Sabag, O., Goel, G., Lale, S., and Hassibi, B. (2021).
\newblock Regret-optimal controller for the full-information problem.
\newblock In \emph{2021 American Control Conference (ACC)}, 4777--4782. IEEE.

\bibitem[{Sabag et~al.(2022)Sabag, Lale, and Hassibi}]{sabag2022optimal}
Sabag, O., Lale, S., and Hassibi, B. (2022).
\newblock Optimal competitive-ratio control.
\newblock \emph{arXiv preprint arXiv:2206.01782}.

\bibitem[{Wang et~al.(2019)Wang, Matni, and Doyle}]{wang2019system}
Wang, Y.S., Matni, N., and Doyle, J.C. (2019).
\newblock A system-level approach to controller synthesis.
\newblock \emph{IEEE Transactions on Automatic Control}, 64(10), 4079--4093.

\bibitem[{Yin et~al.(2021)Yin, Seiler, Jin, and Arcak}]{yin2021imitation}
Yin, H., Seiler, P., Jin, M., and Arcak, M. (2021).
\newblock Imitation learning with stability and safety guarantees.
\newblock \emph{IEEE Control Systems Letters}, 6, 409--414.

\end{thebibliography}

\end{document}